\documentclass[aps,prd,preprint,titlepage,nofootinbib,superscriptaddress]{revtex4-2}

\usepackage[T1]{fontenc}
\usepackage{amsmath,amssymb,amsthm}
\usepackage{mathrsfs}
\usepackage{dsfont}
\usepackage{mathtools}
\usepackage{bm}
\usepackage{graphicx}
\usepackage[hidelinks]{hyperref}

\newcommand{\VV}{\mathcal V}
\newcommand{\LLreg}{\mathcal L}
\newcommand{\jj}{\jmath}
\newcommand{\ct}{\mathrm{ct}}

\begin{document}

\title{Gravitational Origin of Black Hole Thermodynamic Geometry from Boundary Conditions}

\author{Ling-Zheng Xia}
\email{xialingzheng@mail.dlut.edu.cn}
\affiliation{Institute of Theoretical Physics,
Dalian University of Technology, Dalian 116024, China}

\author{Lixin Xu}
\email{lxxu@dlut.edu.cn}
\affiliation{Institute of Theoretical Physics,
Dalian University of Technology, Dalian 116024, China}

\date{\today}

\begin{abstract}
In black hole thermodynamics, Hessian metrics change both their signature and their
singularities from one statistical ensemble to another.  A negative eigenvalue may
therefore signal either a physical instability or only a Legendre transformation.  We
derive this geometry from the bulk action and boundary terms of four-dimensional
Euclidean Einstein--Maxwell--anti-de Sitter gravity.  In the bulk, the radial symplectic
flux pairs the mass and charge with their asymptotic sources, and horizon regularity
selects the equilibrium states.  At infinity, the boundary terms determine the ensemble
polarization by specifying which reservoir parameters are held constant and which
conjugate quantities fluctuate.  Fixing the temperature and electric potential gives the
Ruppeiner metric, whereas adding the Maxwell boundary term fixes the charge and reverses
the corresponding metric sign.  Tracking the boundary conditions thus separates true
stability boundaries from singularities of the Legendre transform.  Local holographic
counterterms leave the metric unchanged, while mixed electric boundary conditions
interpolate between the two ensembles and shift their stability limits.
\end{abstract}

\maketitle

\section{Introduction}\label{sec:intro}

Thermodynamic geometry encodes equilibrium fluctuations in a metric on the space of
states.  In Einstein fluctuation theory, expanding the entropy to second order around
equilibrium gives a Gaussian distribution for the extensive variables.  The negative
entropy Hessian is the inverse covariance of these fluctuations.  Ruppeiner introduced it
as a Riemannian metric on locally stable states~\cite{ruppeiner1979,ruppeiner1995},
conformal to Weinhold's energy geometry~\cite{weinhold1975}.  This line element measures
the
distinguishability of neighboring states against thermal noise and coincides with the
Fisher information metric of the equilibrium distribution~\cite{crooks2007}.  For a charged
system with temperature $T$, entropy $S$, electric potential $\Phi$, and charge $Q$, the
Landau--Lifshitz fluctuation formula~\cite{landau1980} gives the line element
$(dT\,dS+d\Phi\,dQ)/T$.  This quadratic form pairs each intensive source differential with
its conjugate extensive differential, treating both mass and charge as fluctuating
quantities at fixed $T$ and $\Phi$.  The fluctuation interpretation of the entropy Hessian
therefore belongs to the grand canonical ensemble, where its zeros and sign changes locate
black hole spinodals and phase
transitions~\cite{aman2003,liu2010,mansoori2014,dolan2015,mansoori2015,weiliu2019,weiliu2022}.

Changing the statistical ensemble alters both the Hessian metric and its
singularities~\cite{liu2010}.  For example, in asymptotically flat Reissner--Nordstr\"om
black holes, the Ruppeiner metric becomes Lorentzian~\cite{aman2003}, reflecting
instability at fixed potential despite a positive heat capacity at constant charge.
At the same time, a partial Legendre transform from $\Phi$ to $Q$ reverses the electric
second derivative~\cite{callen1985}, leaving the canonical Hessian indefinite although
the system is locally stable.  Without a rule that ties the geometry to the statistical
reservoir, the metric signature alone cannot distinguish a physical instability from a
change of variables.  This ambiguity has motivated Legendre-invariant constructions on
the space of equilibrium states~\cite{quevedo2007,quevedo2008}.

Contact and symplectic thermodynamics clarify why the equations of state alone do not fix
a metric~\cite{mrugala1978,mrugala1990,mrugala1996,bravetti2015,bravetti2019}.  In the
phase space spanned by extensive charges and intensive potentials, equilibrium macrostates
form a Lagrangian submanifold that carries no intrinsic line element.  Defining a Hessian
geometry on this manifold requires a para-complex
polarization~\cite{cruceanu1996,etayo2001,shima2007} that splits the variables into fixed
reservoir sources and fluctuating conjugate responses.  In contact geometry, Legendre
transforms act as changes of contact polarization, and Hessian metrics on the equilibrium
manifold depend on this polarization~\cite{lopezmonsalvo2021,borsboomvisser2026}.  Selecting
a definite
polarization still requires specifying how the physical system couples to its reservoir.

In gravity, the statistical ensemble is selected by the boundary value problem.  In the
Euclidean path integral, a well-posed variational principle requires holding a definite set
of asymptotic field data fixed~\cite{gibbonshawking1977}.  For a charged black hole,
Dirichlet boundary conditions hold the thermal period and electric holonomy constant to
define the grand canonical ensemble~\cite{braden1990,chamblin1999}.  Adding the
Hawking--Ross Maxwell term instead fixes the electric charge and switches to the canonical
ensemble~\cite{hawkingross1995,chamblin1999}.  Anti-de Sitter (AdS) asymptotics
confine thermal radiation and stabilize large black holes in both
settings~\cite{hawkingpage1983,chamblin1999}, much as a finite cavity does in flat
space~\cite{york1986,braden1990}.  Holographic renormalization makes the corresponding
on-shell actions
finite~\cite{deboer2000,skenderis2002,papadimitriouskenderis2005,papadimitriou2010}, while
covariant phase space~\cite{lee1990,iyer1994,hajian2016,hollandswald2013} and spherically
symmetric Hamiltonian
reduction~\cite{kuchar1994,kastrupthiemann1994,loukowhiting1995,loukowintershilt1996}
relate the asymptotic boundary data~\cite{kijowski1979,cattaneo2014} to the conserved
charges.  Together with horizon
regularity~\cite{banados1994,carlipteitelboim1995}, these boundary conditions generate the
thermodynamic potentials and their Legendre transforms~\cite{ma2026}.  However, how
holographic counterterms, mixed boundary conditions, and Legendre degeneracies act on the
induced Hessian metric and its source coordinates remains to be determined.

In this paper, we address these questions for four-dimensional Einstein--Maxwell--AdS
(EM--AdS$_4$) black holes by identifying the asymptotic boundary conditions with a
para-complex polarization of the covariant phase space.  In the bulk, the radial
symplectic flux pairs the conserved charges directly with the proper thermal period and
electric holonomy on the four-dimensional phase space.  Imposing horizon regularity then
selects the Lagrangian submanifold of equilibrium states and fixes the Bekenstein--Hawking
entropy.  At infinity, we show that finite local counterterms on a spherical boundary only
shift the reference energy and leave the Ruppeiner metric invariant.  By contrast,
Legendre boundary terms reverse the metric block of the constrained charge.  Tracking the
fixed sources then distinguishes physical source spinodals, where a static zero mode
appears, from harmless response caustics.  A capacitive Robin boundary condition further
interpolates the spinodals between the two standard ensembles and produces criticality
when the capacitance is smaller than the AdS radius.

This paper is organized as follows.  Section~\ref{sec:thermo-geometry} reviews
para-complex polarizations on thermodynamic phase space and shows how ensemble switches
flip the metric signature.  In Sec.~\ref{sec:zero-modes}, we construct the symplectic
phase space and the equilibrium submanifold from the bulk action and horizon regularity.
Section~\ref{sec:regular-polarization} derives the grand canonical, canonical, and
microcanonical polarizations from asymptotic boundary terms and analyzes counterterm
shears.  Section~\ref{sec:examples} applies the framework to Schwarzschild--AdS$_4$ and
Reissner--Nordstr\"om--AdS$_4$ black holes, including mixed electric boundary conditions,
and Sec.~\ref{sec:discussion} concludes.

\section{Thermodynamic geometry and ensemble polarization}
\label{sec:thermo-geometry}

We first separate the intrinsic equilibrium structure of a thermodynamic system from the
polarization selected by a statistical ensemble.  Consider a system with $n$ conserved
extensive charges $X^a$ and $n$ conjugate intensive
parameters $\lambda_a$ spanning a $2n$-dimensional phase space $\VV$ ($a=1,\dots,n$).
Introducing a bare one-form $\Theta_0$ and exterior derivative $\delta$ on $\VV$ defines
the symplectic two-form
$\Omega=\delta\Theta_0$~\cite{mrugala1978,mrugala1990,mrugala1996,bravetti2015,bravetti2019}:
\begin{equation}
\label{eq:introOmega}
\Omega=\delta X^a\wedge\delta\lambda_a.
\end{equation}
Equilibrium states form an $n$-dimensional submanifold $\LLreg\subset\VV$ with inclusion
$\jj:\LLreg\hookrightarrow\VV$ and exterior derivative $d=\jj^*\delta$.  In
phenomenological thermodynamics, specifying an entropy function $S(X)$ fixes $\LLreg$
through the first law, equations of state, and Maxwell reciprocity,
\begin{equation}
\label{eq:entropy-graph}
dS=\lambda_a\,dX^a,
\qquad
\lambda_a=\partial_a S(X),
\qquad
S_{ab}=S_{ba},
\end{equation}
with $S_{ab}=\partial_a\partial_b S$.  Pulling Eq.~\eqref{eq:introOmega} back to $\LLreg$
then yields the Lagrangian condition
\begin{equation}
\label{eq:Lagrangian-condition}
\jj^*\Omega=dX^a\wedge d\lambda_a=0.
\end{equation}
The phase space and equilibrium manifold are therefore fixed independently of the
statistical ensemble.

Selecting a statistical ensemble adds a boundary functional $I_{\mathrm{bdy}}$ on $\VV$
that shifts the Liouville one-form while preserving $\Omega$:
\begin{equation}
\label{eq:Theta-shift}
\Theta=\Theta_0+\delta I_{\mathrm{bdy}},
\qquad
\delta\Theta=\Omega.
\end{equation}
An ensemble fixes $n$ boundary sources and allows the $n$ conjugate responses to
fluctuate, so $\Theta$ pairs the responses with the differentials of the fixed sources.
In these Darboux coordinates, variations at fixed sources span the subspace where the
source differentials vanish, while variations at fixed responses define a transverse
complement.  Both subspaces are Lagrangian, and this splitting is encoded by a
para-complex structure $\mathcal K$ on $T\VV$ with $\mathcal K^2=\mathds{1}$ compatible
with $\Omega$~\cite{cruceanu1996,etayo2001,shima2007}:
\begin{equation}
\label{eq:K-compatibility}
\Omega(\mathcal K u,v)+\Omega(u,\mathcal K v)=0.
\end{equation}
Variations at fixed sources and at fixed responses carry opposite eigenvalues $-1$ and
$+1$ under $\mathcal K$.  This sign reversal cancels the antisymmetry of $\Omega$ via
Eq.~\eqref{eq:K-compatibility} and yields a flat symmetric neutral metric of signature
$(n,n)$ on $\VV$ with null eigenspaces,
\begin{equation}
\label{eq:eta-general}
\eta(u,v)=\Omega(u,\mathcal K v),
\end{equation}
and pulling $\eta$ back along $\jj$ induces the metric on the embedded equilibrium
manifold,
\begin{equation}
\label{eq:g-pullback}
g=-\frac12\jj^*\eta.
\end{equation}
Pulling $\Theta$ back to $\LLreg$ defines the equilibrium potential $I=-\ln Z$ and
identifies the metric tensor $g$ with its negative source Hessian:
\begin{equation}
\label{eq:general-Hessian}
dI=\jj^*\Theta,
\qquad
g=-\operatorname{Hess}I.
\end{equation}
Because ordinary second derivatives in Eq.~\eqref{eq:general-Hessian} are taken with
respect to these Darboux sources, nonlinear changes of the source coordinates alter
both the polarization and the induced metric.  The Hessian matrix $\mathcal D$ of $I$ is
the Jacobian of the conjugate responses with respect to the fixed sources.  In local
coordinates, $g$ has components $-\mathcal D$ in the source chart and $-\mathcal D^{-1}$
in the response chart.

In the grand canonical ensemble, the reservoir fixes all intensive parameters $\lambda_a$
and allows all extensive charges $X^a$ to fluctuate.  With symmetrized products
$\alpha\gamma=(1/2)(\alpha\otimes\gamma+\gamma\otimes\alpha)$, the Liouville form
$\Theta_{\mathrm{gc}}=X^a\,\delta\lambda_a$ gives
$\eta_{\mathrm{gc}}=2\,\delta X^a\,\delta\lambda_a$, and Eq.~\eqref{eq:g-pullback}
reproduces Ruppeiner's metric $g_R$~\cite{ruppeiner1979}:
\begin{equation}
\label{eq:mainresult}
g_{\mathrm{gc}}=-dX^a\,d\lambda_a
=-S_{ab}\,dX^a dX^b.
\end{equation}
At fixed intensive sources, expanding the off-shell potential
$I_*(X,\lambda)=\lambda_a X^a-S(X)$ around equilibrium to second order in charge
fluctuations $\Delta X$ gives the Gaussian weight
$w\propto\exp[-g_{\mathrm{gc}}(\Delta X,\Delta X)/2]$~\cite{landau1980}.  Its covariance
matrix is therefore $-\mathcal D_{\mathrm{gc}}$~\cite{mrugala1990,crooks2007}, and a
normalizable distribution requires $g_{\mathrm{gc}}$ to be positive definite, or
equivalently the entropy to be strictly concave.

Changing the ensemble adds a Legendre boundary term in Eq.~\eqref{eq:Theta-shift},
modifies the polarization, and induces a distinct metric on the same equilibrium manifold.
Split the indices into $a=(i,k)$, where the reservoir fixes the intensive sources
$\lambda_i$ ($i=1,\dots,n-m$) together with the extensive charges $X^k$
($k=n-m+1,\dots,n$).  Subtracting $\delta(\lambda_k X^k)$ from $\Theta_{\mathrm{gc}}$
shifts the Liouville one-form and equilibrium potential to
\begin{equation}
\label{eq:partial-Legendre-Theta}
\Theta=X^i\,\delta\lambda_i-\lambda_k\,\delta X^k,
\qquad
I=\lambda_i X^i-S.
\end{equation}
Because $\Theta$ pairs the fixed sources $(\lambda_i,X^k)$ with the conjugate responses
$(X^i,-\lambda_k)$, $\mathcal K$ flips sign on each $(X^k,\lambda_k)$ plane and reverses
the corresponding terms in $\eta$:
\begin{equation}
\label{eq:partial-Legendre-eta}
\eta=2\,\delta X^i\,\delta\lambda_i-2\,\delta X^k\,\delta\lambda_k.
\end{equation}
On the equilibrium manifold, Maxwell reciprocity $S_{ik}=S_{ki}$ cancels the off-diagonal
cross terms and reverses the flipped block~\cite{callen1985}:
\begin{equation}
\label{eq:partial-Legendre-g}
g=-dX^i\,d\lambda_i+dX^k\,d\lambda_k
=-S_{ij}\,dX^i dX^j+S_{kl}\,dX^k dX^l.
\end{equation}
For strictly concave entropy, the unconstrained block $-S_{ij}$ is positive definite while
the constrained block $S_{kl}$ is negative definite, so fixing $m$ charges flips $m$
eigenvalues of $g_{\mathrm{gc}}$.  Restricting $g$ to constant $X^k$ projects out the
constrained block and recovers the fluctuation metric of the remaining charges $X^i$.
Fixing all $n$ charges in the microcanonical ensemble reverses the entire metric to
$g_{\mathrm{mc}}=-g_{\mathrm{gc}}$.

Equation~\eqref{eq:partial-Legendre-g} also shows why metric degeneracy alone cannot
identify an instability.  Because $g$ splits diagonally, $\det g=0$ whenever either
$\det S_{ij}=0$ or $\det S_{kl}=0$.  More generally, on any smooth chart of $\LLreg$,
$\det g$ is proportional to the product of the response and source Jacobian determinants,
while $\det\mathcal D$ equals their ratio.  Metric degeneracy therefore occurs if either
map loses rank, but $\det\mathcal D$ separates the two cases.  At a source caustic, or
spinodal ($\det S_{ij}=0$), the source map loses rank and fails to parameterize $\LLreg$
locally, so $\det\mathcal D$ and the susceptibilities diverge~\cite{arnold1985}.  At a
response caustic ($\det S_{kl}=0$), the source coordinates remain regular and only the
response map loses rank, leaving $\mathcal D$ finite with $\det\mathcal D=0$ and no
instability.

\section{Thermodynamic phase space and the equilibrium submanifold}
\label{sec:zero-modes}

We now construct the phase space $(\VV,\Omega)$ and the equilibrium submanifold
$\LLreg$ of Sec.~\ref{sec:thermo-geometry} from the bulk action and horizon regularity
of EM--AdS$_4$ gravity.  Both structures arise in the bulk and remain independent of the
cutoff radius and the asymptotic boundary action.

\subsection{Action, field equations, and the exterior solution family}

We work in geometric units $G=1$ with $\Lambda=-3/\ell^2$.  On the Euclidean contour of
Appendix~\ref{app:intermediate}, the action on a region $\mathcal M$ with radial boundary
$\Sigma_r$ is
\begin{equation}
\label{eq:action}
I_E=I_{\mathrm{bulk}}+I_{\mathrm{bdy}},
\qquad
I_{\mathrm{bulk}}=-\frac{1}{16\pi}\int_{\mathcal M}d^4x\sqrt g
\left(R-2\Lambda+F_{\mu\nu}F^{\mu\nu}\right),
\end{equation}
with $F_{\mu\nu}=\partial_\mu A_\nu-\partial_\nu A_\mu$ and boundary term
$I_{\mathrm{bdy}}$ on $\Sigma_r$.
Stationarity of $I_{\mathrm{bulk}}$ under compactly supported variations gives the
Euclidean Einstein--Maxwell equations
\begin{equation}
\label{eq:Maxwell-eom}
\begin{aligned}
\nabla_\mu F^{\mu\nu}&=0,\\
G_{\mu\nu}+\Lambda g_{\mu\nu}
&=-2F_{\mu\rho}F_\nu{}^\rho+\frac12g_{\mu\nu}F^2.
\end{aligned}
\end{equation}
On the exterior domain $f(r)>0$, we consider the static spherical ansatz
\begin{equation}
\label{eq:ansatz}
ds^2=e^{2\psi(r)}f(r)\,d\tau^2+f(r)^{-1}dr^2+r^2d\Omega_2^2,
\qquad
A=a(r)\,d\tau,\qquad \tau\sim\tau+1,
\end{equation}
with $d\Omega_2^2=d\theta^2+\sin^2\theta\,d\phi^2$.  Normalizing the coordinate
period of $S_\tau^1$ to unity places the physical thermal scale entirely in the
metric profile.  In Appendix~\ref{app:intermediate}, Gauss's law fixes the
electric flux $Q$ and the stress identity $T^\tau{}_\tau=T^r{}_r$ forces $\psi'=0$.
The remaining radial Einstein equation then integrates to the exterior family
\begin{equation}
\label{eq:four-parameter-family}
f(r)=1-\frac{2M}{r}+\frac{Q^2}{r^2}+\frac{r^2}{\ell^2},
\qquad
a(r)=a_0-\frac{NQ}{r},
\qquad
e^{\psi(r)}=N,
\end{equation}
labeled by four integration constants $(M,Q,N,a_0)$.  Outside the outer horizon
$r=r_+>0$, defined as the largest root $f(r_+)=0$, these four constants vary
independently and span the four-dimensional phase space $\VV$.

At $r\to\infty$, the representative conformal boundary metric
$\gamma^{(0)}=\lim_{r\to\infty}(\ell/r)^2 ds^2=N^2 d\tau^2+\ell^2 d\Omega_2^2$ and
gauge potential $A^{(0)}=\lim_{r\to\infty}A=a_0\,d\tau$ identify $(N,a_0)$ with the
proper thermal period $\beta$ and electric holonomy $\mathcal A$ around
$S_\tau^1$:
\begin{equation}
\label{eq:sources}
\beta=\oint_{S_\tau^1}ds_{(0)}=N,
\qquad
\mathcal A=\oint_{S_\tau^1}A^{(0)}=a_0.
\end{equation}
In this conformal frame with sphere radius $\ell$, the proper length $\beta$ of $S_\tau^1$
is diffeomorphism invariant, while gauge transformations vanishing at infinity preserve
$\mathcal A$.

\subsection{Radial symplectic flux and the two invariances of \texorpdfstring{$\Omega$}{Omega}}

Let $\delta$ denote the exterior derivative on the four-dimensional solution space
$\VV$, with $h_{\mu\nu}=\delta g_{\mu\nu}$ and $h=g^{\mu\nu}h_{\mu\nu}$.  Varying
$I_{\mathrm{bulk}}$ in Eq.~\eqref{eq:action} on shell reduces $\delta I_{\mathrm{bulk}}$
to a boundary integral of the Lee--Wald symplectic potential
current~\cite{lee1990,wald1993,iyer1994},
\begin{equation}
\label{eq:theta-g}
\theta_g^{\mu}(\delta g)
=-\frac{\sqrt g}{16\pi}
\left(\nabla_\nu h^{\mu\nu}-\nabla^\mu h\right),
\qquad
\theta_A^{\mu}(\delta A)
=-\frac{\sqrt g}{4\pi}\,F^{\mu\nu}\delta A_\nu.
\end{equation}
Let $\bm\theta$ denote the spacetime three-form associated with
$\theta_g^r+\theta_A^r$.  On the cutoff slice $\Sigma_r\simeq S_\tau^1\times S^2$, the
surface term $\delta I_{\mathrm{bulk}}\big|_{\Sigma_r}=\int_{\Sigma_r}\bm\theta$ defines
the bare radial one-form $\Theta_0=\Theta_{\Sigma_r}$ of Sec.~\ref{sec:thermo-geometry}.
Evaluating this integral on the family~\eqref{eq:four-parameter-family} as in
Appendix~\ref{app:intermediate} gives
\begin{equation}
\label{eq:integrated-theta}
\Theta_{\Sigma_r}
=-\frac12\beta\,\delta M
+\frac12\left(M+\frac{Q^2}{r}+\frac{r^3}{\ell^2}\right)\delta\beta
-Q\,\delta\mathcal A
+\frac{\beta Q}{r}\,\delta Q.
\end{equation}
At fixed radius, the $r$-dependent terms in Eq.~\eqref{eq:integrated-theta} are
$\delta$-exact on $\VV$.  By Eq.~\eqref{eq:Theta-shift}, taking $\delta$ removes them
together with any boundary functional added on the closed slice
$\Sigma_r$~\cite{iyer1994,comperemarolf2008,harlowwu2020}, leaving the same
symplectic form
\begin{equation}
\label{eq:Omega}
\Omega
=\delta\Theta_{\Sigma_r}
=\delta M\wedge\delta\beta-\delta Q\wedge\delta\mathcal A.
\end{equation}
Equation~\eqref{eq:Omega} is therefore independent of both the boundary action and the
cutoff radius $r>r_+$, consistent with on-shell closure $d\bm\omega=0$ of the symplectic
current $\bm\omega=\delta\bm\theta$.

To relate the integration constants $(M,Q)$ in Eq.~\eqref{eq:four-parameter-family} to the
conserved charges at fixed sources, Appendix~\ref{app:intermediate} evaluates the
Iyer--Wald surface integrals on $S_r^2$ for proper time translation $(\xi_\beta,\chi_\beta)$
and gauge shift $(0,1)$.  At any finite radius, these surface integrals yield
\begin{equation}
\label{eq:physical-charges}
H_{\xi_\beta,\chi_\beta}=M,\qquad H_{0,1}=Q.
\end{equation}
Equations~\eqref{eq:sources}, \eqref{eq:Omega}, and~\eqref{eq:physical-charges} thus
identify $(M,Q)$ with the conserved extensive charges conjugate to
$\lambda_a=(\beta,-\mathcal A)$ in the Darboux form~\eqref{eq:introOmega}.  Pairing the
Iyer--Wald charges with the proper period $\beta$ and holonomy $\mathcal A$ directly
selects the natural source parameters of the Boltzmann weight
$\exp(-\beta M+\mathcal A Q)$.  Evaluating the symplectic flux on radial cylinders
complements the spacelike ADM
reduction~\cite{kuchar1994,loukowhiting1995,loukowintershilt1996} and connects the bulk
charges directly to the boundary action.

\subsection{Smooth fillings, horizon flux, and the Lagrangian property}
\label{sec:stokes-generator}

In gravity, the Lagrangian condition~\eqref{eq:Lagrangian-condition} follows directly from
horizon regularity.  A globally regular Euclidean black hole closes smoothly on a compact
filling $\mathcal M_r\simeq D^2\times S^2$ bounded only by $\Sigma_r$ at $r>r_+$ with no
interior boundary.  For the submanifold
$\LLreg\subset\VV$ of such smooth fillings, tangent variations $\delta_1,\delta_2$ along
$\LLreg$ are regular throughout the interior.  Since $\bm\omega$ is closed on shell,
Stokes' theorem gives
\begin{equation}\label{eq:stokes-regular}
(\jj^*\Omega)(\delta_1,\delta_2)
=\int_{\Sigma_r}\bm\omega(\delta_1,\delta_2)
=\int_{\mathcal M_r}d\bm\omega(\delta_1,\delta_2)=0,
\end{equation}
so bulk regularity enforces Eq.~\eqref{eq:Lagrangian-condition} directly and guarantees
that an equilibrium state function exists on $\LLreg$.

At the bolt $r=r_+$ on the nonextremal branch $f'_+=f'(r_+)>0$, the root condition
$f(r_+)=0$ expresses the mass, temperature $T=f'_+/(4\pi)$, potential, and horizon area in
terms of $(r_+,Q)$:
\begin{equation}
\label{eq:m-S}
\begin{aligned}
M(r_+,Q)&=\frac{r_+}{2}\left(1+\frac{Q^2}{r_+^2}+\frac{r_+^2}{\ell^2}\right),
&T(r_+,Q)&=\frac{1}{4\pi r_+}\left(1-\frac{Q^2}{r_+^2}+\frac{3r_+^2}{\ell^2}\right),\\
\Phi(r_+,Q)&=\frac{Q}{r_+},
&\mathscr A_+(r_+)&=4\pi r_+^2.
\end{aligned}
\end{equation}
Terminating an arbitrary exterior solution in $\VV$ at $r=r_+$ leaves a singular inner
boundary with conical deficit angle $2\pi c$ and bolt holonomy
$a_+$~\cite{liwang2022,wuliuwei2026},
\begin{equation}
\label{eq:defect-variables}
c=1-\beta T,
\qquad
a_+=\mathcal A-\beta\Phi.
\end{equation}
Differentiating $M(r_+,Q)$ in Eq.~\eqref{eq:m-S} gives
$\delta M=(T/4)\,\delta\mathscr A_++\Phi\,\delta Q$ and rewrites the symplectic
form~\eqref{eq:Omega} at the bolt as the inner boundary defect flux
\begin{equation}
\label{eq:Omega-bolt}
\delta M\wedge\delta\beta
-\delta Q\wedge\delta\mathcal A
=\frac14\,\delta c\wedge\delta\mathscr A_+
-\delta Q\wedge\delta a_+.
\end{equation}
At infinity $\Omega$ pairs energy with $\beta$ and charge with $\mathcal A$, while at the
bolt it pairs $\mathscr A_+/4$ with the conical deficit~\cite{carlipteitelboim1995} and
$Q$ with $a_+$.

Closing this inner boundary smoothly in the local disk coordinate
\begin{equation}
\label{eq:bolt-chart}
r=r_++\frac{f'_+}{4}\rho^2
\end{equation}
removes the conical singularity at $\rho=0$ with $a(r_+)=0$, setting $c=a_+=0$ in
Eq.~\eqref{eq:defect-variables}, so $\beta=T^{-1}$ and $\mathcal A=\beta\Phi$.  The
metric and gauge profiles are then smooth in $\rho^2$ (Appendix~\ref{app:intermediate}).
Tangent variations therefore extend smoothly across the bolt, the defect flux in
Eq.~\eqref{eq:Omega-bolt} vanishes, and these regular solutions form the Lagrangian
submanifold $\LLreg\subset\VV$.  On $\LLreg$, the mass differential above reduces to the
first law and identifies the state function guaranteed by Eq.~\eqref{eq:stokes-regular}
with the Bekenstein--Hawking entropy~\cite{wald1993,iyer1994},
\begin{equation}
\label{eq:area-entropy}
S=\frac{\mathscr A_+}{4}=\pi r_+^2,
\qquad
dS=\beta\,dM-\mathcal A\,dQ.
\end{equation}

\section{Boundary terms as ensemble polarizations}
\label{sec:regular-polarization}

With $(\VV,\Omega)$ and $\LLreg$ fixed in the bulk, we now evaluate the ensemble
polarizations and induced metrics of Sec.~\ref{sec:thermo-geometry} for the standard
asymptotic boundary conditions of EM--AdS$_4$ gravity.

\subsection{Grand canonical polarization and the Ruppeiner metric}

At large $r$, the bare one-form $\Theta_{\Sigma_r}$ in Eq.~\eqref{eq:integrated-theta}
diverges as $r^3/\ell^2$ and mixes variations of $(M,Q)$ with $(\beta,\mathcal A)$.  For
the Dirichlet problem, we add the Gibbons--Hawking--York term and local counterterm
\begin{equation}
\label{eq:bdy-Dirichlet}
I_{\mathrm{GHY}}+I_{\ct}
=-\frac{1}{8\pi}\int_{\Sigma_r}d^3x\sqrt h\,K
+\frac{1}{8\pi\ell}\int_{\Sigma_r}d^3x\sqrt h
\left(2+\frac{\ell^2}{2}\mathcal R[h]\right),
\end{equation}
with induced metric determinant $h$, extrinsic curvature $K$, and Ricci scalar
$\mathcal R[h]$ on $\Sigma_r$.  In Appendix~\ref{app:intermediate}, adding
$\delta I_{\mathrm{GHY}}+\delta I_{\ct}$ cancels the $O(r^3)$ and $O(r)$ divergences of
$\Theta_{\Sigma_r}$ along with its $\delta M$ term, while $O(r^{-1})$ pieces vanish as
$r\to\infty$.  Holding the intensive sources $(\beta,-\mathcal A)$ fixed defines
the grand canonical ensemble and yields the renormalized Liouville one-form
\begin{equation}
\label{eq:Theta-ren}
\Theta_{\mathrm{gc}}
=\lim_{r\to\infty}\left(\Theta_{\Sigma_r}
+\delta I_{\mathrm{GHY}}+\delta I_{\ct}\right)
=M\,\delta\beta-Q\,\delta\mathcal A,
\qquad
\delta\Theta_{\mathrm{gc}}=\Omega.
\end{equation}

Because $\mathcal M_r$ has no interior boundary, the variation of
$I_{\mathrm{bulk}}+I_{\mathrm{GHY}}+I_{\ct}$ along $\LLreg$ receives
contributions only from $\Sigma_r$ as $r\to\infty$.  By Eq.~\eqref{eq:general-Hessian},
its regular on-shell value $I_{\mathrm{gc}}$ satisfies
\begin{equation}\label{eq:Igc-variation}
dI_{\mathrm{gc}}=\jj^*\Theta_{\mathrm{gc}}
=M\,d\beta-Q\,d\mathcal A,
\end{equation}
in agreement with the Hamilton--Jacobi boundary
relation~\cite{kijowski1979,deboer2000,cattaneo2014}.  On shell,
$R=-12/\ell^2$ and $F_{\mu\nu}F^{\mu\nu}=2Q^2/r^4$, so integrating
$I_{\mathrm{bulk}}+I_{\mathrm{GHY}}+I_{\ct}$ over $r\geq r_+$ gives
\begin{equation}\label{eq:regular-action-value}
I_{\mathrm{gc}}=\frac12\beta\left(M-\frac{r_+^3}{\ell^2}-\frac{Q^2}{r_+}\right)
=\frac14\beta r_+\left(1-\frac{Q^2}{r_+^2}-\frac{r_+^2}{\ell^2}\right).
\end{equation}
Comparing Eq.~\eqref{eq:regular-action-value} with Eqs.~\eqref{eq:m-S}
and~\eqref{eq:area-entropy} recovers the Gibbons--Hawking Legendre
relation~\cite{gibbonshawking1977,chamblin1999}
\begin{equation}\label{eq:entropy-Legendre}
I_{\mathrm{gc}}=\beta M-\mathcal A Q-S,
\end{equation}
so the asymptotic action $I_{\mathrm{gc}}(\beta,\mathcal A)$ and the horizon entropy
$S(M,Q)$ act as conjugate generators of $\LLreg$.  For arbitrary $(M,Q)$ at fixed
$(\beta,\mathcal A)$, continuing this functional off $\LLreg$ gives the off-shell action
$I_*(M,Q,\beta,\mathcal A)=\beta M-\mathcal A Q-\mathscr A_+/4$
\cite{banados1994,carlipteitelboim1995}.  Stationarity of $I_*$ at fixed
$(\beta,\mathcal A)$ enforces the bolt conditions $c=a_+=0$ in
Eq.~\eqref{eq:defect-variables}, while its $(M,Q)$ Hessian generates the Gaussian
fluctuation metric $g_{\mathrm{gc}}$ in Eq.~\eqref{eq:mainresult}.

In the chart $(M,Q,\beta,\mathcal A)$, Eq.~\eqref{eq:Theta-ren} fixes the grand canonical
polarization and neutral metric
\begin{align}
\label{eq:K-gc}
[\mathcal K_{\mathrm{gc}}]_{(M,Q,\beta,\mathcal A)}
&=\operatorname{diag}(-1,-1,+1,+1),\\
\label{eq:eta}
\eta_{\mathrm{gc}}&=2\,\delta M\,\delta\beta-2\,\delta Q\,\delta\mathcal A.
\end{align}
By Eqs.~\eqref{eq:mainresult} and~\eqref{eq:area-entropy}, the induced metric on $\LLreg$
is the Ruppeiner metric $g_R=-\operatorname{Hess}_{(M,Q)}S$~\cite{ruppeiner1979},
\begin{equation}
\label{eq:Ruppeiner}
g_{\mathrm{gc}}
=-dM\,d\beta+dQ\,d\mathcal A
=\frac1T\left(dT\,dS+d\Phi\,dQ\right),
\end{equation}
conformal to Weinhold's energy metric by a factor of
$T$~\cite{weinhold1975,ruppeiner1995,aman2003}.  In the extensive chart $(M,Q)$ with heat
capacity $C_Q=T(\partial S/\partial T)_Q$ and $-S_{MM}=1/(T^2C_Q)$,
Eq.~\eqref{eq:Ruppeiner} and its determinant read
\begin{equation}
\label{eq:ggc-diagonal}
\begin{aligned}
g_{\mathrm{gc}}
&=\frac{dM^2}{T^2C_Q}-2S_{MQ}\,dM\,dQ-S_{QQ}\,dQ^2,\\
\det{}_{(M,Q)}g_{\mathrm{gc}}
&=-\frac{S_{QQ}}{T^2C_Q}-S_{MQ}^2.
\end{aligned}
\end{equation}
Local stability of energy and charge fluctuations at fixed temperature and potential
requires $C_Q>0$ and $\det_{(M,Q)}g_{\mathrm{gc}}>0$, which also ensures $-S_{QQ}>0$.

When $g_{\mathrm{gc}}$ is nondegenerate, inverting the source Jacobian
$\partial\lambda_a/\partial X^b=-(g_{\mathrm{gc}})_{ab}$ gives the grand canonical
response matrix
\begin{equation}
\label{eq:response}
\mathcal D_{\mathrm{gc}}^{ab}
=\frac{\partial X^a}{\partial\lambda_b}
=\frac{\partial(M,Q)}{\partial(\beta,-\mathcal A)}.
\end{equation}
In the intensive chart $(\beta,-\mathcal A)$, Eq.~\eqref{eq:general-Hessian}
writes the metric and the fluctuation covariance as
\begin{align}
\label{eq:gR-lnZ}
&g_{\mathrm{gc}}
=-\mathcal D_{\mathrm{gc}}^{ab}\,d\lambda_a\,d\lambda_b
=-\operatorname{Hess}_\lambda I_{\mathrm{gc}},\\
\label{eq:covariance}
&\langle\Delta X^a\Delta X^b\rangle_{\beta,\mathcal A}
=(g_{\mathrm{gc}}^{-1})^{ab}=-\mathcal D_{\mathrm{gc}}^{ab},
\end{align}
where $(g_{\mathrm{gc}}^{-1})^{ab}$ denotes the inverse metric in the charge basis $X^a$.
At a grand canonical spinodal, $g_{\mathrm{gc}}$ degenerates and the susceptibility
$\mathcal D_{\mathrm{gc}}^{ab}$ diverges.

\subsection{Canonical and microcanonical polarizations}
\label{sec:ensemble-hessians}

On the exterior family~\eqref{eq:four-parameter-family}, the asymptotic Maxwell
variation contributes $-Q\,\delta\mathcal A$ to $\Theta_{\mathrm{gc}}$.  Fixing the
canonical sources $(\beta,Q)$ implements the partial Legendre
transform~\eqref{eq:partial-Legendre-Theta} on the electric plane through the
Hawking--Ross boundary term~\cite{braden1990,hawkingross1995,chamblin1999}
\begin{equation}
\label{eq:IQ}
I_Q=\lim_{r\to\infty}\frac{1}{4\pi}\int_{\Sigma_r}d^3x\sqrt h\,n_\mu F^{\mu\nu}A_\nu
=Q\mathcal A.
\end{equation}
Adding $I_Q$ yields the canonical Liouville one-form and action
\begin{align}
\label{eq:canonical-Darboux}
\Theta_{\mathrm{can}}
&=\Theta_{\mathrm{gc}}+\delta I_Q
=M\,\delta\beta+\mathcal A\,\delta Q,
&\delta\Theta_{\mathrm{can}}&=\Omega,\\
\label{eq:Ican}
I_{\mathrm{can}}
&=I_{\mathrm{gc}}+I_Q=\beta M-S,
&dI_{\mathrm{can}}&=M\,d\beta+\mathcal A\,dQ,
\end{align}
with Helmholtz free energy $F=M-TS$ and $I_{\mathrm{can}}=\beta F$.

Stationarity of $I_{\mathrm{can}}$ now holds $(\beta,Q)$ fixed and identifies
$(M,\mathcal A)$ as conjugate responses, assigning eigenvalue $-1$ to
$\partial_{\mathcal A}$ and $+1$ to $\partial_Q$.  By Eq.~\eqref{eq:partial-Legendre-eta},
this exchange on the electric plane flips the corresponding block of
$\mathcal K_{\mathrm{gc}}$ and reverses the electric pairing in $\eta_{\mathrm{can}}$:
\begin{align}
\label{eq:K-can}
[\mathcal K_{\mathrm{can}}]_{(M,Q,\beta,\mathcal A)}
&=\operatorname{diag}(-1,+1,+1,-1),\\
\label{eq:eta-can}
\eta_{\mathrm{can}}
&=2\,\delta M\,\delta\beta+2\,\delta\mathcal A\,\delta Q.
\end{align}
Pulling $\eta_{\mathrm{can}}$ back to $\LLreg$ shifts the induced metric by
\begin{equation}
\label{eq:g-can}
g_{\mathrm{can}}
=-dM\,d\beta-dQ\,d\mathcal A
=g_{\mathrm{gc}}-2\,dQ\,d\mathcal A.
\end{equation}
Applying Eqs.~\eqref{eq:partial-Legendre-g} and~\eqref{eq:ggc-diagonal} diagonalizes
$g_{\mathrm{can}}$ in $(M,Q)$ and gives its determinant:
\begin{equation}\label{eq:general-canonical-signature}
g_{\mathrm{can}}
=\frac{dM^2}{T^2C_Q}+S_{QQ}\,dQ^2,\qquad
\det{}_{(M,Q)}g_{\mathrm{can}}
=\frac{S_{QQ}}{T^2C_Q}.
\end{equation}
When $S_{QQ}<0$, the flipped electric term is negative, so $g_{\mathrm{can}}$ has
indefinite signature $(+,-)$ for $C_Q>0$ and becomes negative definite for $C_Q<0$.  In
the canonical source chart $(\beta,Q)$, away from spinodals,
$g_{\mathrm{can}}=-\operatorname{Hess}_{(\beta,Q)}I_{\mathrm{can}}$ has response matrix
and thermal energy variance
\begin{equation}
\label{eq:gcan-lnZ}
\mathcal D_{\mathrm{can}}=\frac{\partial(M,\mathcal A)}{\partial(\beta,Q)},
\qquad
\langle\Delta M^2\rangle_{\beta,Q}
=-\left(\frac{\partial M}{\partial\beta}\right)_Q
=T^2C_Q.
\end{equation}
Both an infinite heat capacity and $S_{QQ}=0$ make $\det_{(M,Q)}g_{\mathrm{can}}$ vanish
in Eq.~\eqref{eq:general-canonical-signature}, so metric degeneracy alone cannot
distinguish them.  By contrast, $\det\mathcal D_{\mathrm{can}}=T^2C_QS_{QQ}$ diverges at
a canonical spinodal, where the energy variance grows without bound, and equals zero at a
response caustic ($S_{QQ}=0$).  Along constant charge ($dQ=0$), $g_{\mathrm{can}}$ and
$g_{\mathrm{gc}}$ coincide and recover the thermal fluctuation metric of the energy:
\begin{equation}
\label{eq:dq0-restriction}
g_{\mathrm{can}}\big|_{dQ=0}
=\frac{dM^2}{T^2C_Q}
=\frac{dS^2}{C_Q}.
\end{equation}

A second Legendre transform removes the remaining thermal fluctuation.  Subtracting the
asymptotic Brown--York energy term $\beta M$ from $I_{\mathrm{can}}$ fixes both extensive
charges at the boundary, as in the microcanonical path
integral~\cite{brownyork1993,marolf2018}.  This shift gives
$\Theta_{\mathrm{mc}}=-\beta\,\delta M+\mathcal A\,\delta Q$ and $I_{\mathrm{mc}}=-S(M,Q)$,
reproducing the density of states $e^S$.  By Sec.~\ref{sec:thermo-geometry}, flipping
both conjugate planes reverses the polarization to
$\mathcal K_{\mathrm{mc}}=-\mathcal K_{\mathrm{gc}}$ and
$\eta_{\mathrm{mc}}=-\eta_{\mathrm{gc}}$, so
$g_{\mathrm{mc}}=-\operatorname{Hess}_{(M,Q)}I_{\mathrm{mc}}=\operatorname{Hess}_{(M,Q)}S$
has signature $(-,-)$ when the entropy is strictly concave.  Holding $(M,Q)$ fixed
exchanges sources with responses, freezes all fluctuations, and sets
$\mathcal D_{\mathrm{mc}}=-\mathcal D_{\mathrm{gc}}^{-1}$.  Although
$g_{\mathrm{mc}}=-g_{\mathrm{gc}}$ has the same degeneracy curves and curvature
singularities as $g_{\mathrm{gc}}$, inverting $\mathcal D_{\mathrm{gc}}$ turns each
divergence of $\det\mathcal D_{\mathrm{gc}}$ into a zero of
$\det\mathcal D_{\mathrm{mc}}$.  Every grand canonical spinodal thus becomes a harmless
microcanonical response caustic.
Table~\ref{tab:polarizations} summarizes all three polarizations.

\begin{table}[tb]
\caption{Grand canonical, canonical, and microcanonical polarizations on the common
radial symplectic space $(\VV,\Omega)$ and regular equilibrium submanifold $\LLreg$.
Polarization matrices are written in the chart $(M,Q,\beta,\mathcal A)$, and the
RN--AdS$_4$ discriminants $\Delta_\Phi$ and $\Delta_Q$ are defined in
Sec.~\ref{sec:examples}.}
\label{tab:polarizations}
\footnotesize
\begin{ruledtabular}
\begin{tabular}{lccc}
Quantity & Grand canonical & Canonical & Microcanonical \\
\hline
Boundary action $I_{\mathrm{bdy}}$
& $I_{\mathrm{GHY}}+I_{\ct}$
& $I_{\mathrm{GHY}}+I_{\ct}+I_Q$
& $I_{\mathrm{GHY}}+I_{\ct}+I_Q-\beta M$ \\
Liouville one-form $\Theta$
& $M\,\delta\beta-Q\,\delta\mathcal A$
& $M\,\delta\beta+\mathcal A\,\delta Q$
& $-\beta\,\delta M+\mathcal A\,\delta Q$ \\
On-shell action $I$
& $\beta M-\mathcal A Q-S$
& $\beta M-S$
& $-S$ \\
Sources / Responses
& $(\beta,-\mathcal A)$ / $(M,Q)$
& $(\beta,Q)$ / $(M,\mathcal A)$
& $(M,Q)$ / $(-\beta,\mathcal A)$ \\
Polarization $\mathcal K$
& $\operatorname{diag}(-1,-1,+1,+1)$
& $\operatorname{diag}(-1,+1,+1,-1)$
& $\operatorname{diag}(+1,+1,-1,-1)$ \\
Induced metric $g$
& $-dM\,d\beta+dQ\,d\mathcal A$
& $-dM\,d\beta-dQ\,d\mathcal A$
& $dM\,d\beta-dQ\,d\mathcal A$ \\
Response matrix $\mathcal D$
& $\partial(M,Q)/\partial(\beta,-\mathcal A)$
& $\partial(M,\mathcal A)/\partial(\beta,Q)$
& $\partial(-\beta,\mathcal A)/\partial(M,Q)$ \\
Fluctuating charges
& $(M,Q)$
& $M$
& None \\
Signature ($S$ concave)
& $(+,+)$
& $(+,-)$
& $(-,-)$ \\
RN--AdS$_4$ zero of $\det g$
& $\Delta_\Phi$ (spinodal)
& $\Delta_Q$ (spinodal)
& $\Delta_\Phi$ (response) \\
\end{tabular}
\end{ruledtabular}
\end{table}

\subsection{Counterterms and canonical shears}
\label{sec:choice-boundary}

Legendre boundary terms switch the ensemble by exchanging sources and responses.  Within a
fixed ensemble, adding a finite boundary functional of the fixed sources redefines the
conjugate responses by a canonical shear~\cite{papadimitriou2010}.  For the Dirichlet
polarization with intensive sources $\lambda_a$ and extensive responses $X^a$, adding
$W(\lambda)$ shifts $\Theta_{\mathrm{gc}}$ to
$\Theta_W=\Theta_{\mathrm{gc}}+\delta W$ and transforms the coordinates as
\begin{equation}
\label{eq:shear-definition}
X'^a=X^a+\frac{\partial W(\lambda)}{\partial\lambda_a},\qquad
\lambda'_a=\lambda_a,\qquad
\mathcal S^{ab}=\frac{\partial^2 W(\lambda)}{\partial\lambda_a\partial\lambda_b}
=\mathcal S^{ba}.
\end{equation}
In the coordinate basis $(\partial_X,\partial_\lambda)$, this canonical shear transforms
the polarization and neutral metric to
\begin{equation}
\label{eq:shear-K}
\mathcal K_W=\begin{pmatrix}-\mathds{1}&-2\mathcal S\\0&\mathds{1}\end{pmatrix},\qquad
\eta_W-\eta_{\mathrm{gc}}=2\,\mathcal S^{ab}\,\delta\lambda_a\,\delta\lambda_b.
\end{equation}
The sheared operator still obeys $\mathcal K_W^2=\mathds{1}$ with eigenvalues $\pm 1$,
altering only the $+1$ eigenspace of fixed responses, and Eq.~\eqref{eq:shear-K} shifts
the induced metric $g_{\mathrm{gc}}$ by
\begin{equation}
\label{eq:shear-induced}
g_W-g_{\mathrm{gc}}=-\mathcal S^{ab}\,d\lambda_a\,d\lambda_b.
\end{equation}
A linear functional $W(\lambda)$ has $\mathcal S^{ab}=0$, so it only shifts the reference
zero point of the charges and leaves $\mathcal K_{\mathrm{gc}}$, $\eta_{\mathrm{gc}}$,
and $g_{\mathrm{gc}}$ invariant.  A nonlinear $W(\lambda)$ shifts the action to
$I_W=I_{\mathrm{gc}}+W(\lambda)$, while at fixed sources the factor $e^{-W(\lambda)}$
cancels from the normalized Boltzmann distribution of the charges.  The charge covariance
therefore remains $-\mathcal D_{\mathrm{gc}}^{ab}$, whereas
$g_W=-\operatorname{Hess}_\lambda I_W$ picks up the extra term in
Eq.~\eqref{eq:shear-induced} and no longer coincides with the fluctuation metric
$g_{\mathrm{gc}}$.  Because $\lambda'_a=\lambda_a$, this shear preserves the $-1$
eigenspace of fixed sources and the source projection.  Adding a smooth term
$\mathcal S^{ab}(\lambda)$ to $\mathcal D_{\mathrm{gc}}^{ab}$ therefore leaves the
spinodal poles unchanged.

Beyond $I_{\ct}$ in Eq.~\eqref{eq:bdy-Dirichlet}, any finite local counterterm on the
boundary $S_\tau^1\times S^2$ is constructed from the intrinsic boundary fields
$\gamma^{(0)}$ and $A^{(0)}$.  In three boundary dimensions, parity-even invariants of the
intrinsic curvature and tangential field strength $F^{(0)}=dA^{(0)}$ have even scaling
dimension, excluding logarithmic counterterms and Weyl
anomalies~\cite{skenderis2002,papadimitriouskenderis2005}.  Because
$A^{(0)}=\mathcal A\,d\tau$ has vanishing field strength, no local gauge-invariant density
on the boundary depends on $\mathcal A$.  For constant $\beta$, all intrinsic curvature
scalars of $\gamma^{(0)}$ come from the sphere factor and depend only on $\ell$.  Any
symmetric local gauge invariant is therefore constant on the boundary.  Integrating it
with $\sqrt{\gamma^{(0)}}\propto\beta$ gives at most a linear shift $W=c_1\beta$ of the
reference energy with $\mathcal S^{ab}=0$ and leaves $g_{\mathrm{gc}}$ invariant.  By
contrast, moving the spinodals requires boundary conditions that couple the sources to
the conjugate fluxes, changing which combination is held fixed.  Examples include a
finite cavity
wall~\cite{york1986,anderson2008,witten2021,liusantoswiseman2024,borsboomvisser2026} and
the mixed electric Robin condition of Sec.~\ref{sec:mixed-shear}.

\section{Examples}\label{sec:examples}

We now illustrate the induced metrics, signature rule, and spinodal caustics in
four-dimensional Schwarzschild--AdS and RN--AdS black holes.

\subsection{Schwarzschild--AdS in four dimensions}

On the neutral slice $Q=0$ with $x=r_+/\ell$, the grand canonical and canonical
ensembles reduce to the same thermal polarization with $\Omega=\delta M\wedge\delta\beta$.
Equations~\eqref{eq:m-S} and~\eqref{eq:area-entropy} give $M=(1/2)r_+(1+x^2)$,
$S=\pi r_+^2$, $T=(1+3x^2)/(4\pi r_+)$, and thermal response
\begin{equation}
\label{eq:SAdS-response}
\mathcal D=\frac{\partial M}{\partial\beta}
=\frac{(1+3x^2)^3}{8\pi(1-3x^2)}.
\end{equation}
The induced Ruppeiner metric reads
\begin{equation}
\label{eq:SAdS-Ruppeiner}
g_R=-dM\,d\beta=-\mathcal D^{-1}dM^2
=\frac{8\pi(3x^2-1)}{(1+3x^2)^3}\,dM^2.
\end{equation}
This metric is positive definite on the large black hole branch ($x>1/\sqrt3$), vanishes
at the spinodal $x=1/\sqrt3$, and turns negative definite on the small branch
($x<1/\sqrt3$).  On the small branch, the heat capacity and variance $-\mathcal D=T^2C$
in Eq.~\eqref{eq:covariance} turn negative, so the Gaussian weight of
Sec.~\ref{sec:thermo-geometry} is not normalizable and signals local thermal instability.
Thermal AdS$_4$ with topology
$S_\tau^1\times B^3$ forms a separate branch at vanishing mass where $g_R=0$, reflecting
the absence of saddle fluctuations at leading order in $1/G$.  The Hawking--Page
transition at $x=1$~\cite{hawkingpage1983} occurs at the global action crossing between
thermal AdS$_4$ and the large black hole branch.

\subsection{Reissner--Nordstr\"om--AdS in four dimensions}

For charged RN--AdS$_4$ black holes, we parameterize $\LLreg$ by $(r_+,Q)$ and define
\begin{equation}
\label{eq:deltas}
\begin{aligned}
\Delta_T&=\ell^2(r_+^2-Q^2)+3r_+^4,\quad
\Delta_\Phi=\ell^2(Q^2-r_+^2)+3r_+^4,\quad
\Delta_Q=\ell^2(3Q^2-r_+^2)+3r_+^4,
\\
\Xi&=3\ell^4Q^4+\ell^4r_+^4+12\ell^2Q^2r_+^4
+6\ell^2r_+^6+9r_+^8,
\end{aligned}
\end{equation}
so that $T=\Delta_T/(4\pi\ell^2r_+^3)>0$ defines the nonextremal domain $\Delta_T>0$.
In the horizon chart, Eqs.~\eqref{eq:m-S}, \eqref{eq:Ruppeiner}, and~\eqref{eq:g-can}
give
\begin{equation}
\label{eq:RN-induced-metric}
\begin{aligned}
g_{\mathrm{gc}}&=\frac{2\pi}{\Delta_T}
\left(\Delta_Q\,dr_+^2-4\ell^2Qr_+\,dr_+\,dQ
+2\ell^2r_+^2\,dQ^2\right),\\
g_{\mathrm{can}}&=\frac{2\pi}{\Delta_T^2}
\left[\Delta_T\Delta_Q\,dr_+^2+4\ell^2Qr_+\Delta_Q\,dr_+\,dQ
-2\ell^2r_+^2\left(\Delta_T+4\ell^2Q^2\right)dQ^2\right],
\end{aligned}
\end{equation}
where the polarization flip preserves the thermal $dr_+^2$ term at fixed charge and
reverses the sign of the $dQ^2$ coefficient on the nonextremal branch.

\begin{figure}[tb]
\includegraphics[width=0.78\linewidth]{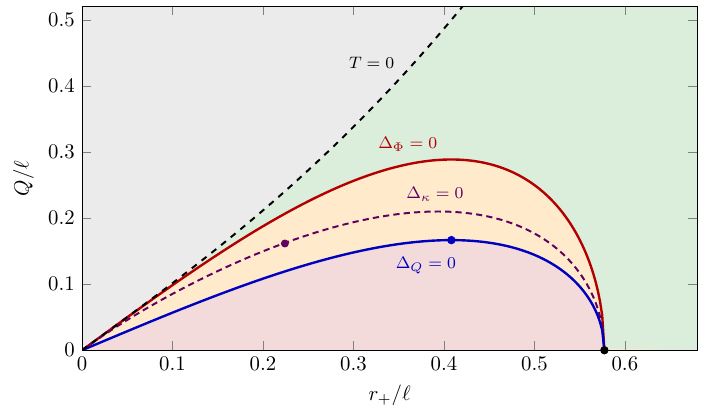}
\caption{Spinodal curves and metric signatures of $(g_{\mathrm{gc}},g_{\mathrm{can}})$ for
RN--AdS$_4$ black holes in the $(r_+/\ell,Q/\ell)$ plane.  Below the extremal curve
$T=0$ (black dashed, with $T<0$ in gray), the grand canonical spinodal $\Delta_\Phi=0$
(red) and canonical spinodal $\Delta_Q=0$ (blue) separate three regions with signatures
$(+,+)/(+,-)$ for $\Delta_\Phi>0$ (green), $(+,-)/(+,-)$ for $\Delta_Q>0>\Delta_\Phi$
(yellow), and $(+,-)/(-,-)$ for $\Delta_Q<0$ (pink).  The purple dashed curve shows the
mixed Robin spinodal $\Delta_\kappa=0$ for $\kappa\ell=2$.  The filled dots mark the
mixed critical point for $\kappa\ell=2$ (purple), the canonical critical point
$(r_c,Q_c)$ (blue), and the neutral spinodal $r_+=\ell/\sqrt{3}$ (black).}
\label{fig:rn-caustics}
\end{figure}

Changing coordinates to the extensive chart $(M,Q)$ via the Jacobian in
Eq.~\eqref{eq:RN-projection-Jacobians} of Appendix~\ref{app:expanded-responses} gives the
metric determinants
\begin{equation}
\label{eq:RN-ensemble-Hessians}
\det{}_{(M,Q)}g_{\mathrm{gc}}
=\frac{32\pi^2\ell^6r_+^6\Delta_\Phi}{\Delta_T^4},
\qquad
\det{}_{(M,Q)}g_{\mathrm{can}}
=-\frac{32\pi^2\ell^6r_+^6\Xi\Delta_Q}{\Delta_T^6}.
\end{equation}
On the nonextremal branch, $\Xi>0$ makes the electric Hessian entry $S_{QQ}$ at fixed
mass negative, while $\Delta_Q$ controls the heat capacity $C_Q$ at fixed charge:
\begin{align}
\label{eq:RN-electric-curvature}
S_{QQ}
&=-\frac{4\pi\ell^2r_+^2\Xi}{\Delta_T^3}<0,\\
\label{eq:Cq}
C_Q
&=2\pi r_+^2\frac{\Delta_T}{\Delta_Q}.
\end{align}
Because $S_{QQ}<0$, $g_{\mathrm{can}}$ has no response caustics, and the two metrics
degenerate only on the spinodals $\Delta_\Phi=0$ and $\Delta_Q=0$ in
Fig.~\ref{fig:rn-caustics}.  Above $\Delta_\Phi=0$, both ensembles are locally stable,
with $g_{\mathrm{gc}}$ positive definite and $g_{\mathrm{can}}$ indefinite due to the
charge constraint.  In the intermediate wedge $\Delta_Q>0>\Delta_\Phi$, fixing the charge
stabilizes the black hole and keeps $\mathcal D_{\mathrm{can}}$ regular across
$\Delta_\Phi=0$, while the grand canonical ensemble becomes unstable.  Both metrics
therefore share the signature $(+,-)$ inside this wedge despite opposite stability,
confirming that metric signature alone cannot determine stability.  Below $\Delta_Q=0$,
$C_Q$ turns negative and both ensembles are unstable.

The curve $\Delta_\Phi=0$ illustrates the contrast between a source spinodal and a
response caustic.  From Appendix~\ref{app:expanded-responses},
$\det\mathcal D_{\mathrm{gc}}\propto\Delta_\Phi^{-1}$ diverges on $\Delta_\Phi=0$, while
$\det\mathcal D_{\mathrm{mc}}=(\det\mathcal D_{\mathrm{gc}})^{-1}\propto\Delta_\Phi$
vanishes there and $\det\mathcal D_{\mathrm{can}}\propto-\Xi/\Delta_Q$ stays finite.  On
the nonextremal branch $\Delta_T>0$, the $(M,Q)$ source Jacobian
$\Delta_T/(2\ell^2r_+^2)$ remains positive, so fixed $(M,Q)$ provide smooth coordinates
across $\Delta_\Phi=0$ and no microcanonical instability arises.  Although
$g_{\mathrm{gc}}$ and $g_{\mathrm{mc}}=-g_{\mathrm{gc}}$ share the same degeneracy curve
and Ricci scalar divergence $R_{\mathrm{mc}}=-R_{\mathrm{gc}}\propto\Delta_\Phi^{-2}$ in
Eq.~\eqref{eq:RN-Ricci-scalars}, only the grand canonical ensemble has a spinodal there.

The two spinodals meet on the neutral slice $Q=0$ at $r_+=\ell/\sqrt3$, recovering
Eq.~\eqref{eq:SAdS-response}.  Along $\Delta_Q=0$ for positive charge, the maximum of $Q$
gives the canonical critical point~\cite{chamblin1999,kubiznakmann2012}
\begin{equation}
\label{eq:rn-critical-point}
r_c=\frac{\ell}{\sqrt6},\qquad
Q_c=\frac{\ell}{6},\qquad
T_c=\frac{\sqrt6}{3\pi\ell}.
\end{equation}
In the asymptotically flat limit $\ell\to\infty$, the canonical spinodal reduces to the
Davies point $r_+=\sqrt3|Q|$~\cite{davies1977}, while $\Delta_\Phi/\ell^2\to Q^2-r_+^2<0$
for all nonextremal black holes.  Flat Reissner--Nordstr\"om black holes are therefore
unstable at fixed potential across the entire nonextremal branch, making $g_{\mathrm{gc}}$
Lorentzian~\cite{aman2003} even on the canonical stable branch $|Q|<r_+<\sqrt3|Q|$ where
$C_Q>0$.  In this limit both Ricci scalars $R_{\mathrm{gc}}$ and $R_{\mathrm{can}}$
approach zero (Appendix~\ref{app:expanded-responses}), while
$\det_{(M,Q)}g_{\mathrm{can}}$ still vanishes at $r_+=\sqrt3|Q|$ and captures the Davies
instability through metric degeneracy.

\subsection{Mixed electric boundary conditions and source shears}
\label{sec:mixed-shear}

Pure source terms in Sec.~\ref{sec:choice-boundary} shear the responses and leave the
spinodals fixed.  Moving the spinodals requires fixing a mixed combination of $\mathcal A$
and $Q$ at the boundary.  In the spherical zero-mode sector, we couple the black hole to a
reservoir of voltage $\Phi_\kappa=\Phi+\kappa Q$ through a series capacitance
$\kappa^{-1}$ with coupling $\kappa\ge 0$.  Subtracting the capacitor boundary term
$W_\kappa(\beta,Q)$ from $I_{\mathrm{gc}}$ defines the mixed Robin action, analogous to a
double-trace deformation~\cite{witten2001,marolf2006}:
\begin{equation}
\label{eq:W-mixed}
I_\kappa=I_{\mathrm{gc}}-W_\kappa(\beta,Q),\qquad
W_\kappa(\beta,Q)=\frac12\kappa\beta Q^2,\qquad \kappa\ge 0.
\end{equation}
Subtracting $\delta W_\kappa$ from $\Theta_{\mathrm{gc}}$ rewrites the Liouville one-form
in the Darboux form
\begin{equation}
\label{eq:shear-mixed-coordinates}
\Theta_\kappa=\Theta_{\mathrm{gc}}-\delta W_\kappa
=M_\kappa\,\delta\beta-Q\,\delta\mathcal A_\kappa,\qquad
(M_\kappa,\mathcal A_\kappa)
=\left(M+\frac12\kappa Q^2,\,\mathcal A+\kappa\beta Q\right).
\end{equation}
This boundary term includes the positive electrostatic energy in $M_\kappa$ and shears the
fixed electric source to the reservoir holonomy $\mathcal A_\kappa=\beta\Phi_\kappa$.
Pulling the resulting mixed polarization back to $\LLreg$ adds a positive electric term to
$g_{\mathrm{gc}}$ that suppresses charge fluctuations:
\begin{equation}
\label{eq:gR-mixed}
g_\kappa
=-dM_\kappa\,d\beta+dQ\,d\mathcal A_\kappa
=g_{\mathrm{gc}}+\frac{\kappa}{T}\,dQ^2.
\end{equation}
In $(M,Q)$, this source shear leaves the thermal line element at fixed charge unchanged
and shifts the electric entry by $\kappa/T$.  Combining Eq.~\eqref{eq:gR-mixed} with
Eq.~\eqref{eq:ggc-diagonal} and evaluating on RN--AdS$_4$ gives the sheared determinant
\begin{equation}
\label{eq:det-gR-mixed}
\det{}_{(M,Q)}g_\kappa
=\det{}_{(M,Q)}g_{\mathrm{gc}}+\frac{\kappa}{T^3C_Q}
=\frac{32\pi^2\ell^6r_+^6\Delta_\kappa}{\Delta_T^4},\qquad
\Delta_\kappa=\Delta_\Phi+\kappa r_+\Delta_Q.
\end{equation}
The spinodal curve $\Delta_\kappa=0$ interpolates continuously between the grand
canonical caustic at $\kappa=0$ and the canonical one as $\kappa r_+\to\infty$.  Inside
the intermediate wedge of Fig.~\ref{fig:rn-caustics}, $\det g_\kappa$ changes sign at
$\kappa_*(r_+,Q)=|\Delta_\Phi|/(r_+\Delta_Q)$ and restores the locally stable signature
$(+,+)$ for $\kappa>\kappa_*$.  A critical point occurs where the small and large black
hole spinodal branches merge at an extremum of the fixed reservoir voltage $\Phi_\kappa$
along $\Delta_\kappa=0$.  As derived in Eq.~\eqref{eq:app-spinodal-Phik-T} of
Appendix~\ref{app:expanded-responses}, this extremum exists only when the capacitance is
smaller than the AdS radius, $\kappa\ell>1$, at the critical horizon radius
\begin{equation}
\label{eq:mixed-critical-point}
r_c(\kappa)=\frac{\sqrt{6\kappa^2\ell^2-2}-2}{6\kappa},\qquad \kappa\ell>1.
\end{equation}
As $\kappa\ell\to 1^+$, $r_c(\kappa)$ and $Q_c(\kappa)$ vanish while the critical
temperature approaches $T_c\to 1/(2\pi\ell)$.  For $\kappa\ell>1$, the critical point
moves outward through $(r_c,Q_c)\simeq(0.224\ell,0.162\ell)$ at $\kappa\ell=2$ in
Fig.~\ref{fig:rn-caustics}, and $T_c(\kappa)$ rises monotonically to the canonical value
in Eq.~\eqref{eq:rn-critical-point}.  In the limit $\kappa r_+\to\infty$, the charge
variance $\langle\Delta Q^2\rangle\simeq T/\kappa$ vanishes while $g_\kappa$ remains
positive definite throughout the canonical stable domain $\Delta_Q>0$.  Because this limit
reaches the same region without flipping the metric sign, the indefinite signature of
$g_{\mathrm{can}}$ there comes solely from the Legendre exchange of $Q$ and $\mathcal A$.

Across the grand canonical, canonical, and mixed ensembles, each source spinodal has a
direct meaning in the Euclidean path integral.  Moving along $\LLreg$ shifts the black
hole mass and charge, and on the spinodal this variation
leaves the fixed boundary sources unchanged to first order.  Each spinodal therefore
supports a regular static zero mode that shifts the bulk charges while preserving the
boundary conditions.  By the stability criterion~\cite{grossperryyaffe1982,reall2001}, this
zero mode marks the onset of a Euclidean negative mode, as confirmed by perturbation
spectra for Schwarzschild--AdS~\cite{prestidge2000} and flat canonical RN black
holes~\cite{monteirosantos2009}.  At a response caustic, by contrast, the source map
remains invertible and no nonzero tangent variation preserves the boundary data, so no
static zero mode arises.

\section{Discussion and conclusion}\label{sec:discussion}

In this paper, we have shown how the boundary value problem of EM--AdS$_4$ gravity
generates black hole thermodynamic geometry.  In the bulk, the radial symplectic flux and
horizon regularity fix the four-dimensional phase space, the equilibrium manifold, and the
Bekenstein--Hawking entropy independently of the ensemble.  The statistical ensemble
enters only at infinity, where the boundary action selects the polarization and induces
the Hessian metric on the equilibrium manifold.  Within this framework, boundary terms
modify the geometry in three ways.  Local counterterms depend only on the fixed sources
and leave the metric and spinodals unchanged.  Robin boundary terms couple the potential
to the charge and move the spinodal curves, while Legendre terms exchange sources with
responses and reverse the corresponding metric block.

This construction clarifies why the same black hole family carries different metrics in
different ensembles.  Switching the ensemble in the Euclidean path integral changes which
quantities are held fixed at infinity and which charges
fluctuate~\cite{mrugala1990,crooks2007}.  Fixing the charge via a Legendre term reverses
the electric block of the metric as a boundary constraint effect, so signature alone
cannot separate stable constrained branches from unstable unconstrained ones.  Similarly,
metric degeneracy and scalar curvature divergences occur on both caustic types (see
Eq.~\eqref{eq:RN-Ricci-scalars}), while physical instabilities arise only at source
spinodals~\cite{mansoori2014,mansoori2015}, where the source projection loses rank and
the susceptibility of the fluctuating charges diverges.  This perspective complements
Legendre-invariant formulations such as
geometrothermodynamics~\cite{quevedo2007,quevedo2008,quevedo2026}.  As emphasized in
recent work~\cite{quevedo2026}, Legendre invariance alone does not eliminate ensemble
dependence or select a unique metric on the equilibrium manifold.  In gravity, the
asymptotic boundary conditions supply that selection rule by fixing the polarization that
governs fluctuations and stability.

The same symplectic construction extends to stationary black holes with a regular
Euclidean section, where horizon regularity and Stokes' theorem define a Lagrangian
equilibrium manifold.  Applying this framework to Kerr--Newman--AdS black holes or a
variable cosmological constant will include angular momentum and thermodynamic volume as
additional conjugate pairs.  Placing the black hole in a finite
cavity~\cite{york1986,anderson2008,witten2021,liusantoswiseman2024,borsboomvisser2026}
also couples the boundary temperature to the mass and shifts the thermal spinodal,
analogous to the electric Robin condition.

\section*{Acknowledgments}
This work was supported by the National Natural Science Foundation of
China under Grant No.~12475047.

\section*{Data availability}
No data were created or analyzed in this study.

\appendix

\section{Bulk variations, conserved charges, and bolt regularity}
\label{app:intermediate}

\subsection{Euclidean electric contour and static field equations}
\label{app:conventions}

In geometric units $G=1$, the Lorentzian Einstein--Maxwell action is
\begin{equation}
\label{eq:Lorentzian-specialized}
S_L=\frac{1}{16\pi}\int d^4x\sqrt{-g}
\left(R-2\Lambda-F_{\mu\nu}F^{\mu\nu}\right).
\end{equation}
Under $t=-i\tau$, a real Lorentzian electrostatic potential $A_L=-a(r)\,dt$ continues
to the imaginary Euclidean one-form $-ia(r)\,d\tau$.  Expressing the Euclidean action
in terms of the real potential $A=a(r)\,d\tau$ reverses the sign of the Maxwell
kinetic term and yields $I_{\mathrm{bulk}}$ in Eq.~\eqref{eq:action}.  This saddle
continuation keeps the on-shell action real and matches the Gibbons--Hawking
evaluation~\cite{gibbonshawking1977,hawkingross1995}.

The bulk equations~\eqref{eq:Maxwell-eom} reduce on the ansatz~\eqref{eq:ansatz} to a
coupled radial system.  The $\tau$ component of the Maxwell equation in
Eq.~\eqref{eq:Maxwell-eom} gives $(r^2 e^{-\psi}a')'=0$ and integrates to Coulomb's law
$r^2 e^{-\psi(r)}a'(r)=Q$ with electric flux constant $Q$.  The Euclidean Maxwell stress
tensor $T_{\mu\nu}$ on the right side of Eq.~\eqref{eq:Maxwell-eom} then has equal mixed
temporal and radial components $T^\tau{}_\tau=T^r{}_r=-Q^2/r^4$.
Subtracting the $\tau$ and $r$ Einstein equations gives
$G^r{}_r-G^\tau{}_\tau=2f(r)\psi'(r)/r=0$ and fixes $e^{\psi(r)}=N>0$ to be constant on
the exterior domain $f(r)>0$.  Integrating the radial equation
$(rf)'-1-3r^2/\ell^2=-Q^2/r^2$ and $a'=NQ/r^2$ then yields the exterior
family~\eqref{eq:four-parameter-family}.

\subsection{Radial Lee--Wald contractions and boundary variations}

On the exterior solution family~\eqref{eq:four-parameter-family}, the nonzero metric
variations at fixed $r$ are
\begin{equation}
\label{eq:metric-variation-details}
\delta f=-\frac{2\,\delta M}{r}+\frac{2Q\,\delta Q}{r^2},
\qquad
h_{rr}=-\frac{\delta f}{f^2},
\qquad
h_{\tau\tau}=2Nf\,\delta N+N^2\delta f.
\end{equation}
The $\delta f$ contributions cancel in the trace $h$, and the
radial covariant divergence evaluates to
\begin{equation}
\label{eq:trace-divergence-details}
h=\frac{2\,\delta N}{N},
\qquad
\nabla_\nu h^{r\nu}
=\frac{2\,\delta M}{r^2}-\frac{f'\delta N}{N},
\qquad
\nabla^r h=0,
\end{equation}
while for the Maxwell field,
\begin{equation}
\label{eq:Maxwell-variation-details}
F^{r\tau}=\frac{Q}{Nr^2},
\qquad
\delta A_\tau=\delta a_0-\frac{Q}{r}\delta N
-\frac{N}{r}\delta Q.
\end{equation}
Substituting Eqs.~\eqref{eq:trace-divergence-details} and~\eqref{eq:Maxwell-variation-details}
into Eq.~\eqref{eq:theta-g} with $\sqrt g=Nr^2\sin\theta$ and integrating over
$\Sigma_r$ cancels the $4\pi$ sphere factor and gives
\begin{align}
\label{eq:Theta-g-details}
\Theta_{\Sigma_r}^{g}
&=\int_{\Sigma_r}d\tau\,d\theta\,d\phi\,\theta_g^r
=-\frac12 N\,\delta M+\frac14 r^2 f'(r)\,\delta N,\\
\label{eq:Theta-A-details}
\Theta_{\Sigma_r}^{A}
&=\int_{\Sigma_r}d\tau\,d\theta\,d\phi\,\theta_A^r
=-Q\,\delta a_0
+\frac{Q^2}{r}\,\delta N
+\frac{NQ}{r}\,\delta Q.
\end{align}
Using Eqs.~\eqref{eq:four-parameter-family} and~\eqref{eq:sources}, the
sum of Eqs.~\eqref{eq:Theta-g-details} and~\eqref{eq:Theta-A-details} gives the bare
radial one-form $\Theta_{\Sigma_r}$ in Eq.~\eqref{eq:integrated-theta}.

Evaluating the Gibbons--Hawking--York term and the counterterm functional $I_{\ct}$ in
Eq.~\eqref{eq:bdy-Dirichlet} on the ansatz~\eqref{eq:ansatz} gives
\begin{equation}
\label{eq:IGHY-Ict}
I_{\mathrm{GHY}}=-\frac12 N\left(\frac{r^2f'}2+2rf\right),
\qquad
I_{\ct}=\frac12 N\sqrt f\left(\frac{2r^2}{\ell}+\ell\right).
\end{equation}
Their variations on $\VV$ read
\begin{equation}
\begin{aligned}
\delta I_{\mathrm{GHY}}
&=-\frac12\left(\frac{r^2f'}2+2rf\right)\delta N
+\frac32 N\,\delta M
-\frac{NQ}{r}\,\delta Q,\\
\delta I_{\ct}
&=\frac12\sqrt f\left(\frac{2r^2}{\ell}+\ell\right)\delta N
+\frac{N}{2\sqrt f}
\left(\frac{2r^2}{\ell}+\ell\right)
\left(-\frac{\delta M}{r}+\frac{Q\,\delta Q}{r^2}\right).
\end{aligned}
\end{equation}
Expanding at large $r$ with $\sqrt f=r/\ell+\ell/(2r)-M\ell/r^2+O(r^{-3})$ shows that
$\Theta_{\Sigma_r}+\delta I_{\mathrm{GHY}}+\delta I_{\ct}$ is free of $O(r^3)$ and
$O(r)$ divergences.  Its $O(1)$ coefficient of $\delta M$ cancels, and the remaining
$O(r^{-1})$ terms vanish as $r\to\infty$ to leave Eq.~\eqref{eq:Theta-ren}.

\subsection{Hamiltonian charges and normalization}
\label{app:charge-normalization}

To establish Eq.~\eqref{eq:physical-charges} for the conserved charges $(M,Q)$, we
continue the exterior family~\eqref{eq:four-parameter-family} to the Lorentzian geometry
\begin{equation}
\label{eq:Lorentzian-charge-fields}
ds_L^2=-N^2f(r)\,dt^2+f(r)^{-1}dr^2+r^2d\Omega_2^2,
\qquad A_L=-a(r)\,dt,
\end{equation}
with volume form $\epsilon_{tr\theta\phi}=Nr^2\sin\theta$ and sphere orientation
$d\theta\wedge d\phi$.  For the action~\eqref{eq:Lorentzian-specialized}, combined
diffeomorphism and gauge transformations act on $A_L$ by
$\delta_{(\xi,\chi)}A_L=\pounds_\xi A_L+d\chi$.  The Iyer--Wald Noether charge
two-forms~\cite{lee1990,wald1993,iyer1994} read
\begin{equation}
\label{eq:Lorentzian-Noether-forms}
(\bm Q^g_\xi)_{\mu\nu}
=-\frac{1}{16\pi}\epsilon_{\mu\nu\rho\sigma}\nabla^\rho\xi^\sigma,
\qquad
\bm Q^A_{\xi,\chi}
=-\frac{1}{4\pi}(\xi\cdot A_L+\chi)\,*F_L,
\end{equation}
and the surface charge variation for a fixed generator is
\begin{equation}
\label{eq:Hamiltonian-variation}
\delta H_{\xi,\chi}
=\int_{S^2_\infty}
\left[\delta(\bm Q^g_\xi+\bm Q^A_{\xi,\chi})
-\iota_\xi\bm\theta_L\right].
\end{equation}

For $\xi=\partial_t$ at fixed $\chi$, the integrals of the Noether forms and the
contractions of the Lorentzian symplectic potential $\bm\theta_L$ over $S^2_r$ are
\begin{equation}
\label{eq:Qg-sphere}
\int_{S^2_r}\bm Q^g_{\partial_t}
=\frac14 Nr^2f',
\qquad
\int_{S^2_r}\bm Q^A_{\partial_t,\chi}
=Q\left(\chi-a_0+\frac{NQ}{r}\right),
\end{equation}
and
\begin{equation}
\label{eq:iota-theta-g}
\int_{S^2_r}\iota_{\partial_t}\bm\theta^g_L
=-\frac12 N\,\delta M+\frac14 r^2f'\,\delta N,
\qquad
\int_{S^2_r}\iota_{\partial_t}\bm\theta^A_L
=-Q\,\delta a(r).
\end{equation}
Subtracting Eq.~\eqref{eq:iota-theta-g} from the variation of Eq.~\eqref{eq:Qg-sphere}
cancels the $\delta N$ and $\delta a(r)$ terms and all radial dependence at finite $r$:
\begin{equation}
\label{eq:delta-H-coordinate}
\delta H_{\partial_t,\chi}
=N\,\delta M+(\chi-a_0)\,\delta Q.
\end{equation}

At fixed $(N,a_0)$, the unit proper time translation generator with respect to the
conformal boundary metric $\gamma^{(0)}$ in the stationary gauge is the combined symmetry
\begin{equation}
\label{eq:combined-symmetry}
(\xi_\beta,\chi_\beta)
=\left(N^{-1}\partial_t,\,\frac{a_0}{N}\right).
\end{equation}
Equation~\eqref{eq:delta-H-coordinate} and the unit gauge generator $(0,1)$ with
$\int_{S^2_r}\bm Q^A_{0,1}=Q$ then give
\begin{equation}
\label{eq:normalized-charge-check}
\delta H_{\xi_\beta,\chi_\beta}=\delta M,
\qquad
\delta H_{0,1}=\delta Q.
\end{equation}
Integrating at fixed sources and fixing the reference constants in the
vacuum limit $M,Q\to 0$ confirms Eq.~\eqref{eq:physical-charges}.

\subsection{Smoothness of tangent variations at the bolt}

In the disk chart~\eqref{eq:bolt-chart} of Sec.~\ref{sec:stokes-generator}, the regular
exterior family~\eqref{eq:four-parameter-family} expands around $\rho=0$ as
\begin{equation}
\label{eq:bolt-expansion}
\begin{aligned}
ds^2&=\left[1-\frac{f''_+}{8}\rho^2+O(\rho^4)\right]d\rho^2
+4\pi^2\rho^2\left[1+\frac{f''_+}{8}\rho^2+O(\rho^4)\right]d\tau^2
+[r_+^2+O(\rho^2)]\,d\Omega_2^2,\\
A&=\frac{\pi Q}{r_+^2}\rho^2d\tau+O(\rho^4)d\tau,
\end{aligned}
\end{equation}
with $f''_+=f''(r_+)$.  In disk Cartesian coordinates
$(x,y)=(\rho\cos 2\pi\tau,\rho\sin 2\pi\tau)$, the radial and angular metric
coefficients are analytic in $\rho^2=x^2+y^2$ and share unit leading value at $\rho=0$.
Because $d\rho^2+4\pi^2\rho^2 d\tau^2=dx^2+dy^2$ and $\rho^2 d\tau=(x\,dy-y\,dx)/(2\pi)$,
the metric and gauge field in Eq.~\eqref{eq:bolt-expansion} are smooth in $(x,y)$ across
$\rho=0$.  Parametric variations along $\LLreg$ differentiate only the smooth
higher-order coefficients and therefore extend smoothly across $\rho=0$ on
$D^2\times S^2$.  Changing the radial coordinate
$r(\rho)$ to the exterior chart shifts $\bm\omega$ by an exact differential that
integrates to zero over $\Sigma_r$, confirming Eq.~\eqref{eq:stokes-regular}.

\section{RN--AdS coordinate Jacobians and mixed electric shear}
\label{app:expanded-responses}

\subsection{Grand canonical and canonical Jacobians and response matrices}

In terms of $(r_+,Q)$, $\Delta_T$, and $\Delta_Q$ from Eq.~\eqref{eq:deltas}, define
the auxiliary polynomials
\begin{equation}
B=\ell^2(Q^2+r_+^2)+3r_+^4,
\qquad
C=\ell^2Q^2+3r_+^4.
\end{equation}
Differentiating Eq.~\eqref{eq:m-S} on $\LLreg$ gives the three nontrivial differentials
\begin{equation}
\label{eq:RN-differentials}
\begin{aligned}
dM&=\frac{\Delta_T}{2\ell^2r_+^2}\,dr_++\frac{Q}{r_+}\,dQ,\\
d\beta&=\frac{4\pi\ell^2r_+^2}{\Delta_T^2}
\left(-\Delta_Q\,dr_++2\ell^2Qr_+\,dQ\right),\\
d\mathcal A&=\frac{4\pi\ell^2r_+}{\Delta_T^2}
\left(-2QC\,dr_++r_+B\,dQ\right).
\end{aligned}
\end{equation}
Using $B\Delta_Q-4\ell^2Q^2C=\Delta_T\Delta_\Phi$ and $\Delta_T B+4\ell^2Q^2C=\Xi$,
pairwise exterior products among $dQ$ and Eq.~\eqref{eq:RN-differentials} yield the four
projection determinants
\begin{equation}
\label{eq:RN-projection-Jacobians}
\begin{aligned}
\det\frac{\partial(M,Q)}{\partial(r_+,Q)}
&=\frac{\Delta_T}{2\ell^2r_+^2},
&
\det\frac{\partial(\beta,-\mathcal A)}{\partial(r_+,Q)}
&=\frac{16\pi^2\ell^4r_+^4\Delta_\Phi}
{\Delta_T^3},
\\
\det\frac{\partial(M,\mathcal A)}{\partial(r_+,Q)}
&=\frac{2\pi\Xi}{\Delta_T^2},
&
\det\frac{\partial(\beta,Q)}{\partial(r_+,Q)}
&=-\frac{4\pi\ell^2r_+^2\Delta_Q}{\Delta_T^2}.
\end{aligned}
\end{equation}
The zeros of the two source determinants locate the spinodals $\Delta_\Phi=0$ and
$\Delta_Q=0$, while combining Eq.~\eqref{eq:RN-induced-metric} with
$\det[\partial(M,Q)/\partial(r_+,Q)]$ yields Eq.~\eqref{eq:RN-ensemble-Hessians}.

Applying the chain rule to Eq.~\eqref{eq:RN-differentials} gives the grand canonical
response matrix $\mathcal D_{\mathrm{gc}}=\partial(M,Q)/\partial(\beta,-\mathcal A)$,
the canonical response matrix
$\mathcal D_{\mathrm{can}}=\partial(M,\mathcal A)/\partial(\beta,Q)$, and their
determinants:
\begin{align}
\label{eq:D11}
\mathcal D_{\mathrm{gc}}
&=-\frac{\Delta_T}{8\pi\ell^4r_+^4\Delta_\Phi}
\begin{pmatrix}
\Xi & 4\ell^2r_+QC \\
4\ell^2r_+QC & 2\ell^2r_+^2\Delta_Q
\end{pmatrix},
&
\det\mathcal D_{\mathrm{gc}}
&=\frac{\Delta_T^4}{32\pi^2\ell^6r_+^6\Delta_\Phi},
\\[6pt]
\label{eq:RN-Dcan}
\mathcal D_{\mathrm{can}}
&=\begin{pmatrix}
-\dfrac{\Delta_T^3}{8\pi\ell^4r_+^4\Delta_Q}
&\dfrac{2QC}{r_+\Delta_Q}\\[8pt]
\dfrac{2QC}{r_+\Delta_Q}
&\dfrac{4\pi\ell^2r_+^2\Delta_\Phi}{\Delta_T\Delta_Q}
\end{pmatrix},
&
\det\mathcal D_{\mathrm{can}}
&=-\frac{\Xi}{2\ell^2r_+^2\Delta_Q}.
\end{align}
The matrix $\mathcal D_{\mathrm{can}}$ extends smoothly across $\Delta_\Phi=0$
whenever $\Delta_Q\neq 0$, and $\det\mathcal D_{\mathrm{can}}$ differs from
$\det_{(M,Q)}g_{\mathrm{can}}$ in Eq.~\eqref{eq:RN-ensemble-Hessians} by the squared
chart Jacobian $(\partial M/\partial\beta)_Q^2=T^4C_Q^2\propto\Delta_Q^{-2}$.

The two-dimensional Ricci scalars of $g_{\mathrm{gc}}$ and $g_{\mathrm{can}}$ in
Eq.~\eqref{eq:RN-induced-metric} take the closed forms
\begin{equation}
\label{eq:RN-Ricci-scalars}
R_{\mathrm{gc}}
=-\frac{9r_+^2C\left(\Delta_\Phi-2r_+^4\right)}{\pi\Delta_T\Delta_\Phi^2},
\qquad
R_{\mathrm{can}}
=\frac{9r_+^2\Delta_T^3P_{\mathrm{can}}}{\pi\Xi^2\Delta_Q^2},
\end{equation}
with $P_{\mathrm{can}}=3\ell^4Q^4-7\ell^4Q^2r_+^2+12\ell^2Q^2r_+^4-5\ell^2r_+^6+9r_+^8$.
On the nonextremal branch $\Delta_T>0$, $R_{\mathrm{mc}}=-R_{\mathrm{gc}}$ and
$R_{\mathrm{gc}}$ diverge as $\Delta_\Phi^{-2}$ on $\Delta_\Phi=0$, while
$R_{\mathrm{can}}$ diverges as $\Delta_Q^{-2}$ on $\Delta_Q=0$.  Away from the spinodals,
both Ricci scalars vanish as $O(\ell^{-2})$ in the asymptotically flat limit
$\ell\to\infty$.

\subsection{Mixed electric polarization shear}
\label{app:mixed-shear}

On $\LLreg$, differentiating $(M_\kappa,\mathcal A_\kappa)$ in
Eq.~\eqref{eq:shear-mixed-coordinates} gives $dM_\kappa=dM+\kappa Q\,dQ$ and
$d\mathcal A_\kappa=d\mathcal A+\kappa\beta\,dQ+\kappa Q\,d\beta$, cancelling the
$\kappa Q\,d\beta\,dQ$ cross terms in $-dM_\kappa\,d\beta+dQ\,d\mathcal A_\kappa$ to yield
Eq.~\eqref{eq:gR-mixed}.  Their exterior products satisfy $dM_\kappa\wedge dQ=dM\wedge dQ$
and $d\beta\wedge(-d\mathcal A_\kappa)=d\beta\wedge(-d\mathcal A)-\kappa\beta\,d\beta\wedge dQ$,
leaving $\det[\partial(M_\kappa,Q)/\partial(r_+,Q)]$ unchanged and giving the mixed
source determinant from Eq.~\eqref{eq:RN-projection-Jacobians}:
\begin{equation}
\label{eq:app-mixed-source-det}
\det\frac{\partial(\beta,-\mathcal A_\kappa)}{\partial(r_+,Q)}
=\det\frac{\partial(\beta,-\mathcal A)}{\partial(r_+,Q)}
-\kappa\beta\det\frac{\partial(\beta,Q)}{\partial(r_+,Q)}
=\frac{16\pi^2\ell^4r_+^4\left(\Delta_\Phi+\kappa r_+\Delta_Q\right)}{\Delta_T^3}.
\end{equation}
Dividing by $\det[\partial(M,Q)/\partial(r_+,Q)]$ reproduces Eq.~\eqref{eq:det-gR-mixed}.
The mixed response matrix
$\mathcal D_\kappa=\partial(M_\kappa,Q)/\partial(\beta,-\mathcal A_\kappa)$ has
determinant $\det\mathcal D_\kappa=\Delta_T^4/(32\pi^2\ell^6r_+^6\Delta_\kappa)$ and
electric susceptibility
$(\partial Q/\partial\mathcal A_\kappa)_\beta=\Delta_T\Delta_Q/(4\pi\ell^2r_+^2\Delta_\kappa)$,
while the heat capacity at fixed $\Phi_\kappa$ diverges on $\Delta_\kappa=0$:
\begin{equation}
\label{eq:app-C-Phik}
C_{\Phi_\kappa}
=T\left(\frac{\partial S}{\partial T}\right)_{\Phi_\kappa}
=2\pi r_+^2\frac{(1+\kappa r_+)\Delta_T}{\Delta_\kappa}.
\end{equation}
Expanding $\Delta_\kappa=(1+\kappa r_+)(3r_+^4-\ell^2r_+^2)+(1+3\kappa r_+)\ell^2Q^2$ and
solving $\Delta_\kappa=0$ for $Q^2$ expresses the reservoir voltage
$\Phi_\kappa=(1+\kappa r_+)Q/r_+$ and temperature $T$ along the spinodal as functions of
$r_+$ alone:
\begin{equation}
\label{eq:app-spinodal-Phik-T}
\Phi_\kappa^2(r_+)
=\frac{(1+\kappa r_+)^3\left(\ell^2-3r_+^2\right)}{\ell^2(1+3\kappa r_+)},
\qquad
T(r_+)
=\frac{\kappa\ell^2+3r_++6\kappa r_+^2}{2\pi\ell^2(1+3\kappa r_+)}.
\end{equation}
Setting $d(\Phi_\kappa^2)/dr_+=0$ gives $6\kappa^2r_+^2+4\kappa r_++1-\kappa^2\ell^2=0$
and yields the positive root $r_c(\kappa)$ in Eq.~\eqref{eq:mixed-critical-point} for
$\kappa\ell>1$.  Substituting $r_c(\kappa)$ back into $\Delta_\kappa(r_c,Q_c)=0$ fixes
$Q_c(\kappa)$ to the left of the $Q$ maximum on $\Delta_\kappa=0$ because criticality
occurs at an extremum of $\Phi_\kappa$ rather than $Q$.  Evaluating $T(r_+)$ in
Eq.~\eqref{eq:app-spinodal-Phik-T} at $r_c(\kappa)$ gives $T_c\to 1/(2\pi\ell)$ as
$\kappa\ell\to 1^+$ and recovers Eq.~\eqref{eq:rn-critical-point} as
$\kappa\ell\to\infty$.

\bibliography{references}

\end{document}